\documentclass[pra,final,twocolumn,showpacs,superscriptaddress,aps]{revtex4-1}
\usepackage{graphicx}
\usepackage{amssymb}
\usepackage{amsmath}
\usepackage{color}
\usepackage{multirow}
\usepackage{afterpage}

\begin{document}
\title{Modulation instability in quasi two-dimensional spin-orbit coupled Bose-Einstein condensates}

\author{S. Bhuvaneswari}
\affiliation{Department of Physics, Bharathidasan University, Palkalaiperur, Tiruchirappalli 620024, India.}

\author{K. Nithyanandan}
\affiliation{Department of Physics, Pondicherry University, Puducherry 605014, Puducherry, India.}

\author{P. Muruganandam}
\affiliation{Department of Physics, Bharathidasan University, Palkalaiperur, Tiruchirappalli 620024, India.}

\author{K. Porsezian}
\affiliation{Department of Physics, Pondicherry University, Puducherry 605014, Puducherry, India.}

\begin{abstract}
We theoretically investigate the dynamics of modulation instability (MI) in two-dimensional spin-orbit coupled Bose-Einstein condensates (BECs). The analysis is performed for equal densities of pseudo-spin components. Different combination of the signs of intra- and inter-component interaction strengths are considered, with a particular emphasize on repulsive interactions. We observe that the unstable modulation builds from originally miscible condensates, depending on the combination of the signs of the intra- and inter-component interaction strengths. The repulsive intra- and inter-component interactions admit instability and the MI immiscibility condition is no longer significant. Influence of interaction parameters such as spin-orbit and Rabi coupling on MI are also investigated. The spin-orbit coupling (SOC) inevitably contributes to instability regardless of the nature of the interaction. In the case of attractive interaction,  SOC manifest in enhancing the MI. Thus, a comprehensive study of MI in two-dimensional spin-orbit coupled binary BECs of pseudo-spin components is presented.
\end{abstract}

\maketitle
\section{Introduction}
\label{sec:1}
Study of spin-orbit (SO) coupled Bose-Einstein condensates (BECs) is one among the important topics of current research in the context of macroscopic quantum phenomena. Spin-orbit coupling (SOC) describes the interaction between the particle's spin and orbital momentum and plays a crucial role for many physical phenomena in condensed matter systems including spin-Hall effect, topological insulators, spintronics and so on~\cite{Nayak2008,Hasan2010,Hasan2012,Konig2007,Bernevig2006,Kane2005}. The synthetic SOC in BECs was experimentally achieved very recently. In this realization techniques, two Raman laser beams were used to couple with two component BECs \cite{Lin2011}. The momentum transfer between laser beams and atoms leads to synthetic SOC~\cite{Ruseckas2005,Zhang2010,Stanescu2007,Campbell2011}. SOC has been realized with cold atomic gases by designating the hyperfine atomic states as pseudo-spins and coupling them with Raman laser beams~\cite{Zhang2012,Pengjun2012, Cheuk2012}. For instance, in the case of $^{87}$Rb the pseudo-spin states are $\lvert \uparrow \rangle =\lvert F=1,m_{F}=0\rangle $ and $\lvert \downarrow \rangle =\lvert F=1,m_{F}=-1\rangle $, which are generated using pair of Raman laser beams.

SO coupled BECs have been studied extensively in different contexts including phase separation, strip phases~\cite{Wang2010}, spotlighting the phase transition~\cite{Li2012}, vortices with or without rotations~\cite{Radic2011}, and so on. In addition, the study of topological excitations, for example, skyrmions, has also attracted much along these directions~\cite{Liu2012}. Further, matter wave solitons such as bright and dark solitons have been studied in quasi-one-dimensional with attractive and repulsive SO coupled BEC \cite{Achilleos2013,Achilleosdark2013}. It should be noted that most of the studies on SO coupled BECs were primarily focused on quasi-one-dimensional systems. Only a few studies were devoted on multi-dimensional SO coupled BECs. However, there were few important studies reported in the context of two-dimensional SO coupled BECs, for instance, the dynamics of vortices, the existence of vortex-antivortex pair, to mention few ~\cite{He2013,SongHe2013,Cheng2014,Salasnich2014,Jin2014}. Recently, the study on two-dimensional SO coupled BEC of mixed Rashba-Dresselhaus type and Rabi couplings have earned particular interest \cite{Dressel,Bychkov}. Thus, it is more appropriate, realistic and interesting to study SO coupled BECs in two- and three-dimensions systems. Particularly, here we emphasize on the study of instability of plane wave in two-dimensional SO coupled BECs, in the framework of MI analysis \cite{He2013,SongHe2013}.

The degree of instability in a BEC can be characterized by the MI. MI is an instability process and identified as a requisite mechanism to understand various physics effects in nonlinear systems. The phenomenon of MI was first observed in hydrodynamics by Benjamin and Feir in 1967~\cite{BF}. In the same year, Ostrovskii predicted the possibility of MI in optics~\cite{Ostrovsky1967} and later explained in detail by Hasegawa et. al. in 1973 in the context of optical fibers~\cite{Hasegawa:73}.  The MI is a general phenomenon occurring in many nonlinear wave equation and is of particular interest in dispersive nonlinear systems. In conventional dispersive nonlinear systems, MI manifest as a result of the constructive interplay between dispersion and nonlinearity. Such that any deviation from the steady state in the form of perturbation leads to an exponential growth of the weak perturbation, resulting in a break-up of the carrier wave into trains of soliton-like pulses \cite{Agrawal2013}.  In addition, MI has been widely studied in various branches such as fluid dynamics~\cite{BF}, magnetism~\cite{Assalauov1985}, plasma physics~\cite{Taniuti1968} and BEC~\cite{Kourakis2005}.

In the context of BEC, the MI has been given considerable importance over a long period of time, owing to its fundamental and applied interest in various aspects. In particular, MI has been found to be relevant in understanding the formation and propagation of solitonic waves~\cite{Kevrekidis}, and also apparent in explaining the domain formation~\cite{Zhang:02} and quantum phase transition~\cite{Smerzi}. MI has been studied extensively in BECs for both single~\cite{Theocharis} and two-component systems~\cite{Kasamatsu2006}, and realized experimentally as well~\cite{Strecker2002}. In the case of single-component BECs, the MI has been found to be feasible only for attractive interaction (self-focusing nonlinearity), in such case, the phase fluctuations caused by MI leads to the formation of soliton trains. However, the breakthrough work by Goldstein and Meystre opens up the possibility of MI even for the repulsive interactions~\cite{Goldstein1997}, in similar lines to the case of cross-phase modulation induced instability in nonlinear optics~\cite{Agrawal:87,Nithyanadnan}.  Thus, the two component BEC finds particular interest in the study of MI, as it helps to achieve instability even in repulsive interactions. In the case of SO coupled BEC, the MI in one-dimensions was recently explored in Ref.~\cite{Bhat2015}, and the higher dimensional case is still an open problem.  Thus, inspired by the special features of SO coupling and the physical relevance of two-component BEC system,  we intend to study the dynamical behavior of MI in SO coupled BEC in two-dimensions. In this paper, we present a systematic study of MI in quasi-two-dimensional SO coupled BECs with the inclusion of Rashba and Dresselhaus SO couplings.  By considering small perturbation approximation, we obtain linearized GP equation. Further, the interplay between dispersion and nonlinear effects have been studied in terms of system parameters. We have also summarized the growth of MI gain for different combinations of intra- and inter-component of interaction strengths in the presence and absence of SO coupling.

The organization of the paper is as follows: After a detailed introduction in Sec.~\ref{sec:1}, the Sec.~\ref{sec:2}  features the theoretical model for the case of SO coupled BECs. In Sec.~\ref{sec:3}, we present the MI dispersion relation through linear stability analysis, and systematically explained the effect of SOC and Rabi coupling for a different combination of inter- and intra-component interactions strength. Sec.~\ref{sec:4}, features the results and discussion followed by conclusion in Sec.~\ref{sec:5}.

\section{Theoretical Model}

\label{sec:2}
We consider the spin-orbit coupled Bose-Einstein condensates confined in a harmonic trap with equal Rashba and Dresselhaus couplings described, within the framework of mean field theory by an energy functional of the following form \cite{Lin2011}
\begin{align}
E=\int _{-\infty }^{+\infty } \varepsilon\, d\tilde{x}\, d\tilde{y},
\end{align}
where,
\begin{align}
\varepsilon = \frac{1}{2} \left(\Psi ^{\dagger } H_{0} \Psi +\tilde{g}_{11}\lvert \psi_{\uparrow} \rvert^{4}+\tilde{g}_{22}\lvert \psi_{\downarrow} \rvert^{4}+2 \tilde{g}_{12} \lvert \psi _{\uparrow} \rvert^{2} \lvert \psi _{\downarrow} \rvert^{2} \right), \label{eq:e2d}
\end{align}
$\Psi =(\psi _{\uparrow}, \; \psi_{\downarrow} )^{T}$ is the condensate wave function, $\psi_{\uparrow} $ and $\psi_{\downarrow} $ are associated with the pseudo-spin components. The model Hamiltonian $H_{0}$ in Eq.~(\ref{eq:e2d}) assumes the form,
\begin{align}
H_{0}=\left[\frac{\hat{p}^{2}}{2m}+ {V}(\tilde r) \right]+\frac{\hbar \lambda  }{2}\sigma _{x}-\frac{\hbar \tilde{k}_{L}}{m}\hat{p}_{\tilde{x}}\sigma _{z} \label{eq:heq2d}
\end{align}
where, $\hat{p}=-\mathrm{i}\hbar (\partial _{\tilde{x}},\;\partial _{\tilde{y}})$ is the momentum operator, ${V}(\tilde r)=\frac{1}{2} m [\omega_{\bot}^{2}(\tilde x^{2}+\tilde y^{2})+\omega _{z}^{2} \tilde z^{2} ]$ is a quasi-2D harmonic trapping potential where $\omega _{z}\gg \omega _{\perp }$, $\lambda  $ is the frequency of Raman coupling, $\sigma_{x,z}$ are Pauli spin matrices and $\tilde{k}_{L}$ is the wave number of the Raman laser which couples the two hyperfine states. The effective two dimensional coupling constant $\tilde{g}_{ij} =4\pi \hbar^{2}  a_{ij}/m$, $(i,j=1, 2)$ represents the intra- $(\tilde{g}_{11}, \tilde{g}_{22})$ and inter- component ($\tilde{g}_{12}$) interaction strengths, which are defined by the corresponding $s$-wave scattering lengths $ a_{ij}$ and atomic mass $m$. Measuring energy in units of the radial trap frequency ($\omega _{\bot}$), i.e., $\hbar \omega _{\bot }$, length in units of harmonic oscillator length, $a_{\bot }=\sqrt{\hbar /(m\omega _{\bot})}$, and time in units of $\omega _{\bot}^{-1}$ the following dimensionless Gross-Pitaevskii (GP) equations can be derived for different components of $\psi_{1,2} $ from Eq.~(\ref{eq:e2d}) as~\cite{Bao2015}
\begin{subequations}
\label{eq:gp2d}
\begin{align}
\mathrm{i}\frac{\partial\psi_{1}} {\partial t}= & \left[ -\frac{1}{2} \nabla_\perp^2
   +V(r)+  g_{11}\lvert \psi _{1} \rvert ^{2}+g_{12} \lvert\psi _{2} \rvert ^{2}  \right] \psi _{1} \notag \\ &
   +\mathrm{i}k_{L}\frac{\partial}{\partial x}  \psi_{1} + \Gamma \psi_{2}, \\
\mathrm{i}\frac{\partial\psi_{2} } {\partial t}= & \left[ -\frac{1}{2} \nabla_\perp^2
    +V(r)+  g_{22} \lvert\psi _{2} \rvert ^{2}+g_{12} \lvert\psi_{1} \rvert^{2} \right]\psi _{2 }\notag \\ &
    -\mathrm{i}k_{L}\frac{\partial}{\partial x}  \psi_{2} + \Gamma \psi_{1},
\end{align}
\end{subequations}%
where, $V(r)=(x^2 + y^2)/2$, $x=\tilde{x}/a_{\bot} $, $y=\tilde{y}/a_{\bot} $, $t=\omega_{\bot } \tilde{t}$, $k_{L}=\tilde{k}_{L}a_{\bot}$, $\Gamma=\lambda /2 \omega_{\perp } $, $g_{ij}=4\pi N a_{ij}/a_{\bot}$  and $\psi_{1,2}=\psi _{\uparrow, \downarrow}a_{\bot}^{3/2} /\sqrt{N}$. In the following, we shall proceed with the study of modulational instability in the above two-dimensional model Eq. (\ref{eq:gp2d}) for spin-orbit coupled BECs.

\section{Analysis of modulation instability}
\label{sec:3}

\subsection{Linear Stability Analysis}
The fundamental framework of MI analysis relies on the linear stability analysis (LSA), such that the steady state solution is perturbed by a small amplitude/phase, and then study whether the perturbation amplitude grows or decays~\cite{Agrawal2013}. For this purpose, we consider a continuous wave (CW) state of the miscible SO coupled BECs with the two-dimensional density $n_{j0}=\lvert\psi _{j0}\rvert^{2}$ of the form
\begin{align}
\psi_{j}(x,y,t)=\sqrt{n_{j0}}~\mathrm{e}^{-\mathrm{i}\mu t}.
\end{align}
Then the stability of the SO coupled BECs can be examined by assuming the perturbed wave functions as
\begin{align}
\label{perturbation}
\psi_{j}(x,y,t)=(\sqrt{n_{j0}}+\delta \phi_{j})~\mathrm{e}^{-\mathrm{i}\mu t},
\end{align}
A set of linearized equations for the perturbation can be obtained by using Eq.~(\ref{perturbation}) in Eq.~(\ref{eq:gp2d})
\begin{subequations}\label{eq:gpe2d}
\begin{align}
\mathrm{i}\frac{\partial (\delta \phi_{1} )} {\partial t}&=  -\frac{1}{2} \left(\frac{\partial^{2} (\delta \phi_{1} )} {\partial x^{2}} + \frac{\partial^{2} (\delta \phi_{1} )} {\partial y^{2}}\right) +\mathrm{i}k_{L}\frac{\partial(\delta \phi_{1} )}{\partial x} \notag \\ &
 + \Gamma \left(\delta \phi_{2} -\sqrt{\frac{n_{20}}{n_{10}}} (\delta \phi_{1} )\right) +g_{11}n_{10}\left(\delta \phi_{1} +\delta \phi_{1} ^{*}\right)\notag \\ &
+g_{12}\sqrt{n_{10}n_{20}}\left(\delta \phi_{2} +\delta \phi_{2} ^{*}\right), \\
\mathrm{i}\frac{\partial (\delta \phi_{2} )} {\partial t}&=  -\frac{1}{2} \left(\frac{\partial^{2} (\delta \phi_{2} )} {\partial x^{2}} + \frac{\partial^{2} (\delta \phi_{2} )} {\partial y^{2}}\right) -\mathrm{i}k_{L}\frac{\partial(\delta \phi_{2} )}{\partial x} \notag \\ &
 + \Gamma \left(\delta \phi_{1} -\sqrt{\frac{n_{10}}{n_{20}}} (\delta \phi_{2} )\right) +g_{22}n_{20}\left(\delta \phi_{2} +\delta \phi_{2} ^{*}\right)\notag \\ &
+g_{12}\sqrt{n_{10}n_{20}}\left(\delta \phi_{1} +\delta \phi_{1} ^{*}\right) ,
\end{align}
\end{subequations}
where the symbol $^*$ denotes complex conjugate.
Assuming a general solution of the form
\begin{align}\label{eq:delphi}
\delta \phi _{j}=\zeta _{j} \cos\left(k_{x} x+k_{y} y-\Omega t\right)+\mathrm{i} \eta _{j} \sin\left(k_{x} x+k_{y} y-\Omega t\right),
\end{align}
where $k_x$ and $k_y$, are the wavenumbers and $\zeta _{j}$ and $\eta _{j}$ $(j=1,2)$ are the amplitudes of wavefunction, and $\Omega $ is the eigenfrequency. We further assume that two pseudo-spin states of equal density $n_{10}=n_{20}=n$. A straightforward substitution of Eq.~(\ref{eq:delphi}) in Eq.~(\ref{eq:gpe2d}) yields the following dispersion relation for $\Omega $.
\begin{align}
& \Omega ^{4}- \Omega ^{2}\left[\frac{1}{4} \left(K-2 \Gamma \right)\left(2K+G_{1}+G_{2}\right)+2k_{x}^{2}k_{L}^{2}+2\Gamma G_{12}\right]\notag \\
& +\frac{\Omega}{2} k_{x}k_{L}(K-2\Gamma )(G_{1}-G_{2}) \notag \\
& +K\left[k_{L} ^{2}\left(k_{x}^{2}k_{L} ^{2}+2\Gamma G_{12}-\frac{1}{4} \left(K-2 \Gamma \right)\left(2K+G_{1}+G_{2}\right)\right)\right.\notag \\
&  \quad\quad \left.+\left(\frac{K}{4}-\Gamma \right)\left(\frac{1}{4}(K+G_{1})(K+G_{2})-G_{12}^{2}\right)\right]  \notag \\
& -\frac{1}{2} k_L^2 k_y^2 \bigg[\frac{2\Gamma -K}{2}(G_1+G_2-4\Gamma )+2 k_L^2 k_x^2-K^2 \notag \\
& \quad\quad +4\Gamma  \left(G_{12}+K-\Gamma \right)\bigg] = 0,
\label{eq:des2d}
\end{align}
where $K= k_{x}^{2}+k_{y}^{2}$ and $G_1 =4 g_{11} n-2\Gamma$, $G_2=4 g_{22} n-2\Gamma$, and $G_{12}=2 g_{12} n+\Gamma $ are the modified intra- and inter-components interaction strengths, respectively.
For equal strengths of intra-component interactions, i.e., $a_{11} = a_{22}$  ($g_{11} = g_{22} = g$), the dispersion relation recast into a simpler form as
\begin{align}
\Omega _{\pm}^{2}=\frac{1}{2}\left(\Lambda \pm\sqrt{\Lambda ^{2}+4\Delta }\right), \label{eq:om2d}
\end{align}
with
\begin{subequations}
\begin{align}
\Lambda  = &\, \frac{1}{2} \left(K-2 \Gamma \right)\left(K+G\right)+2k_{x}^{2}k_{L} ^{2}+2\Gamma G_{12},
\end{align}
\begin{align}
\Delta = &\, \frac{1}{2} k_L^2 k_y^2 \bigg[(2\Gamma -K)(G-2\Gamma )+2 k_L^2 k_x^2-K^2 \notag \\
& \quad\quad +4\Gamma  \left(G_{12}+K-\Gamma \right)\bigg]\notag \\
& -K\left[k_{L} ^{2}\left(k_{x}^{2}k_{L} ^{2}+2\Gamma G_{12}-\frac{1}{2} \left(K-2 \Gamma \right)\left(K+G\right)\right)\right. \notag \\
&  \quad\quad \left.+\left(\frac{K}{4}-\Gamma \right)\left(\frac{1}{4}(K+G)^{2}-G_{12}^{2}\right)\right],
\end{align}
\end{subequations}
where, $G_1 =G_2 =G$.
The above Eq.~(\ref{eq:om2d}) is the dispersion relation corresponding to the stability of the miscible SO coupled BECs. As it is known from the theory of MI, the system exhibit stable configuration for all real values of $k_{x}$ and $k_{y}$, if $\Omega _{\pm}^{2}$ is positive $(\Omega _{\pm}^{2}>0)$. If $\varLambda >0$, the eigenfrequency $\Omega _{+}$ is always real but $\Omega _{-}$ may be real or imaginary, which is dependent on $\Delta $. If the eigenfrequency $\Omega _{\pm}$ has an imaginary part, the spatially modulated perturbation become exponential with time, as it is obvious from the form of $\delta \phi_{j}$. On the other hand, for negative value of $\varLambda $ $(\varLambda <0)$, $\Omega _{\pm}^{2}$ need not to be positive. In such case, $\Omega _{\pm}^{2}$ is characterized by the values of $\Delta$. For $\varLambda >0$ the value of lower branch $\Omega _{-}^{2}$ is negative if, $\Delta <0$. Similarly for $\varLambda >0$ the value of upper branch $\Omega _{+}^{2}$ is negative when  $\Delta <0$. Regardless of anything $\Omega _{-}^{2}$ is always negative, and therefore, the MI sets in via the exponential growth of the weak perturbations. The MI growth rate is defined as $\xi \equiv \lvert \text{Im} (\Omega _{\pm})\rvert$.  Following the mathematical calculation pertaining to the dispersion relation corresponding to the stability/instability of the system, the subsequent sections are dedicated to the study on the effect of SOC in the MI.

\subsection{Effect of Rabi coupling in the MI of SO coupled BECs}

In order to study the effect of Rabi coupling in the MI, we turn off the SOC by making $k_{L} = 0$.  For better insight, we consider two special cases, (i) one without Rabi coupling $(\Gamma = 0)$ and (ii) other in the presence of Rabi coupling $(\Gamma \neq 0)$.

\subsubsection{Zero Rabi coupling}

In absence of Rabi coupling ($\Gamma = 0$), the eigenfrequency of the system for $k_x\neq k_y$ assumes the form,
\begin{align}
 \Omega _{\pm}^{2}=\frac{1}{2}\left[K\left(K+2 n (g\pm g_{12})\right)\right] \label{omega_sol1}
\end{align}
One can infer from the above Eq.~(\ref{omega_sol1}), that based on the sign/nature of the interaction strength, the $\Omega _{\pm}$  may be real or imaginary.  It is obvious from the combination of signs of intra and inter-component interactions, $\Omega _{+}$ is found to be real in the following  cases:
\begin{enumerate}
\item[(i)] both intra- (g) and  inter- (g$_{12}$) component interactions are repulsive,
\item[(ii)] attractive intra-component and repulsive inter-component interactions, and
\item[(iii)] repulsive intra-component and attractive inter-component interactions.
\end{enumerate}
For attractive intra- and inter-component interactions $\Omega _{+}$ becomes imaginary and thereby inevitably contributes to MI. However, $\Omega _{-}$ becomes imaginary for all cases. Thus, as far as MI is concern, $\Omega _{-}$ contribute better to the instability in all means than the $\Omega _{+}$ counterpart. It is worth mentioning, at $k_y=0$, our results completely agree with the Ref.~\cite{Bhat2015}, and could reproduce the results of the MI in the conventional two-component system as in Ref.~\cite{Kasamatsu2006}.

\subsubsection{Non-zero Rabi coupling}

Next, we study the effect of Rabi coupling on MI  by considering any finite value for $\Gamma$ ($\Gamma\neq0$). Here the dispersion relation as given by Eq.~(\ref{eq:om2d}) can be modified as follows
\begin{align}
 \Omega _{+}^{2}=\frac{1}{2}\left[K\left(K+2 n (g\pm g_{12})\right)\right]. \label{omega_sol2},
\end{align}
with $k_x\neq k_y$, and in order to highlight the effect of Rabi coupling, the coefficient of SOC is turned off, i.e., $k_{L} =0$.  It is straightforward to notice that $\Omega_{+}^{2}$ given by Eq.~(\ref{omega_sol2}) for $\Gamma = 0$ is similar to  Eq.~(\ref{omega_sol1}) for the case of zero Rabi coupling. Therefore, the instability/stability condition as defined by the zero Rabi coupling in the earlier section is completely applicable here as well. Hence, for non-zero Rabi coupling, $\Omega_{+}^{2}$  is not different from that of zero Rabi coupling, which implies that $\Omega_{+}^{2}$  is independent of $\Gamma$. On the other hand, $\Omega_{-}^{2} $ is found to be significantly influenced by Rabi coupling and can be expressed as
\begin{align}
\Omega _{-}^{2}=\frac{K^{2}}{4} + \left(Kn-4\Gamma \right) \left(g-g_{12}\right) + 2\Gamma\left(2\Gamma-K\right).
\label{omega_sol3}
\end{align}
The effect of Rabi coupling from Eq.~(\ref{omega_sol3}) can be better explored for three representative cases of $\Gamma$, namely (i) $\Gamma = 0$, (ii) $\Gamma > 0$, and (iii) $\Gamma < 0$. For $\Gamma = 0$, Eq.~(\ref{omega_sol3}) reverts to the expression for $\Omega _{-}$ as given by Eq.~(\ref{omega_sol1}), and therefore, will not be discussed again here. Our particular focus is on  $\Gamma = 0$ and $\Gamma > 0$.  For $\Gamma < 0$, $\Omega_{-}$ is imaginary only for repulsive intra- and attractive inter- component interaction, and for all other cases $\Omega_{-}$ does not contribute to MI, as it is real. However, the effect of Rabi coupling is more pronounced for $\Gamma > 0$, as the instability/stability conditions qualitatively differ from the previous cases.  It is found that the $\Omega_{-}$ is unstable for all combination interactions, except the repulsive intra- and attractive inter-component interaction. Perhaps, the magnitude of intra- and inter-component interaction strengths are rather identified to be deterministic for MI. It is observed that for repulsive intra- and inter-component interactions the instability is possible only when $\lvert g \rvert >\lvert g_{12} \rvert$. Similarly, for attractive interaction, the condition for MI can be modified as $\lvert g_{12} \rvert >\lvert g \rvert$.

\begin{figure}[!ht]%
\centering\includegraphics[width=0.99\linewidth]{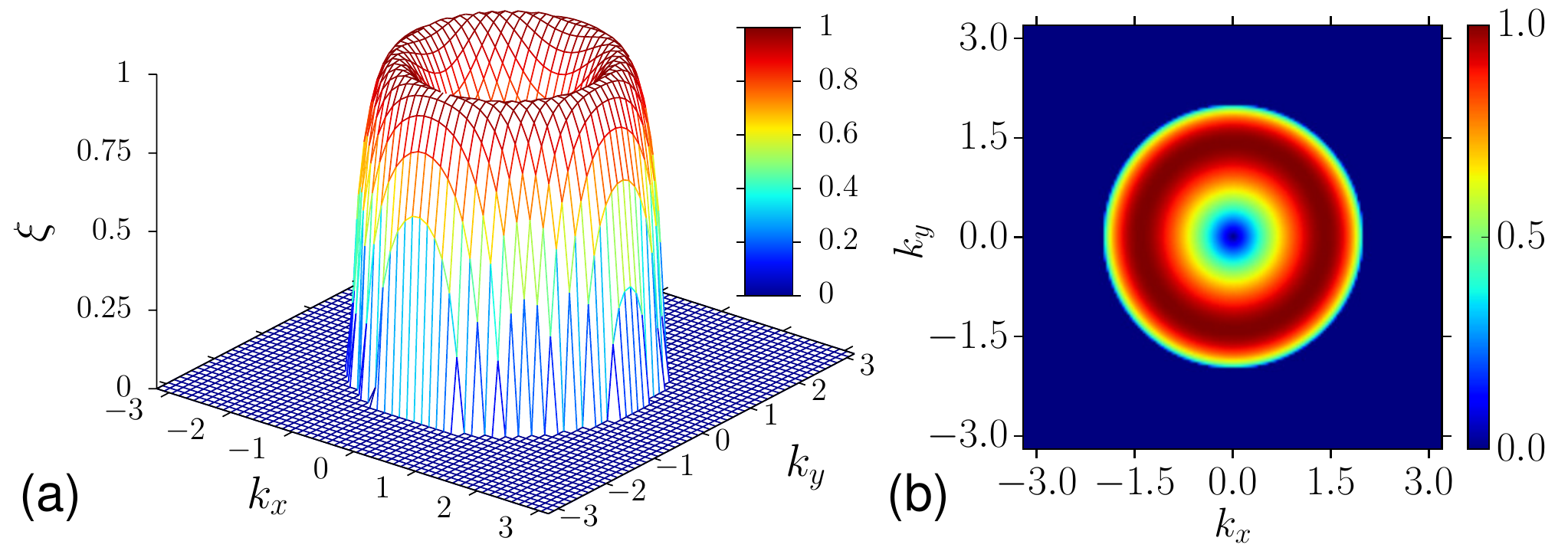}
\caption{(color online) (a) Three-dimensional (3D) surface plot showing the MI gain, $\xi = \lvert \text{Im} (\Omega _{-})\rvert$, and (b) the corresponding two-dimensional (2D) contour plot for the parameters $k_{L}=0$, $\Gamma =1$, $n=1$, $g=2$ and $g_{12}=1$.}
\label{fig-ls-1}
\end{figure}

For a better understanding of the effect of Rabi coupling, as a representative case, we have shown in Fig.~\ref{fig-ls-1}, the MI gain corresponding to the repulsive intra- and inter-component interactions with  $k_{L}=0$, $\Gamma =1$, $g=2$, $g_{12}=1$ and $n=1$. It should be noted, the condition for instability ($\lvert g \rvert >\lvert g_{12} \rvert$) in repulsive interactions is true for the above choice of parameters.
It is evident, from the existence of instability region, that the MI is caused by Rabi coupling for repulsive intra- and inter-component interactions. It should be noted that the instability region is symmetric in momentum space on either side of the  wave numbers, $k_x$ and $k_y$.

Overall, it is apparent from the above discussion on the influence of Rabi coupling in the instability that out of the different choices of Rabi coupling strengths, the condition $\Gamma>1$ is found to carry more information about the MI. Therefore, in the subsequent section we shall study the effect of SO coupling by fixing the Rabi coupling strength as $\Gamma=1$.

\section{The effect of Rabi and spin-orbit coupling}
\label{sec:4}

One can draw out a conclusion from the previous section, that the sign/nature of the interaction significantly influences the stability/instability of the system. For better insight, in the following section, we would like to briefly emphasize the effect of different combinations of intra- and inter- component interaction strength with the inclusion of both Rabi ($\Gamma \neq 0$) and SOC~($k_L \neq 0$). We consider following four representative cases to study MI in the SO coupled BEC system.
\begin{enumerate}
\item[A.] Both repulsive intra- and inter- component interactions ($g > 0$,  $g_{12} > 0$).
\item[B.] Attractive intra- and repulsive inter- component interactions ($g < 0$, $g_{12} > 0$).
\item[C.] Repulsive intra- and attractive inter- component interactions ($g > 0$,  $g_{12} < 0$).
\item[D.] Both attractive intra- and inter- component interactions ($g < 0$,  $g_{12} < 0$).
\end{enumerate}

\subsection{Repulsive intra- and inter-component interactions}
\label{sec:4A}
Here, we consider self repulsive intra- $(g>0)$ and repulsive inter- $(g_{12}>0)$ components of modified interactions $G_1, G_2 $ and $G_{12}$. Our investigation follows from the general dispersion relation for non-zero SO and Rabi coupling as given by Eq.~(\ref{eq:om2d}).  It is apparent from Eq.~(\ref{eq:om2d}), the expression $\Omega_{\pm}$ can be real or complex depends on the sign of $\varLambda >0$ and $\varLambda^2 +4\Delta$. For $\varLambda >0$, the upper branch $\Omega_{+}$ will be imaginary only for $\varLambda^2 +4\Delta < 0$, and therefore contribute to MI.
\begin{figure}[!ht]
\centering\includegraphics[width=0.99\linewidth]{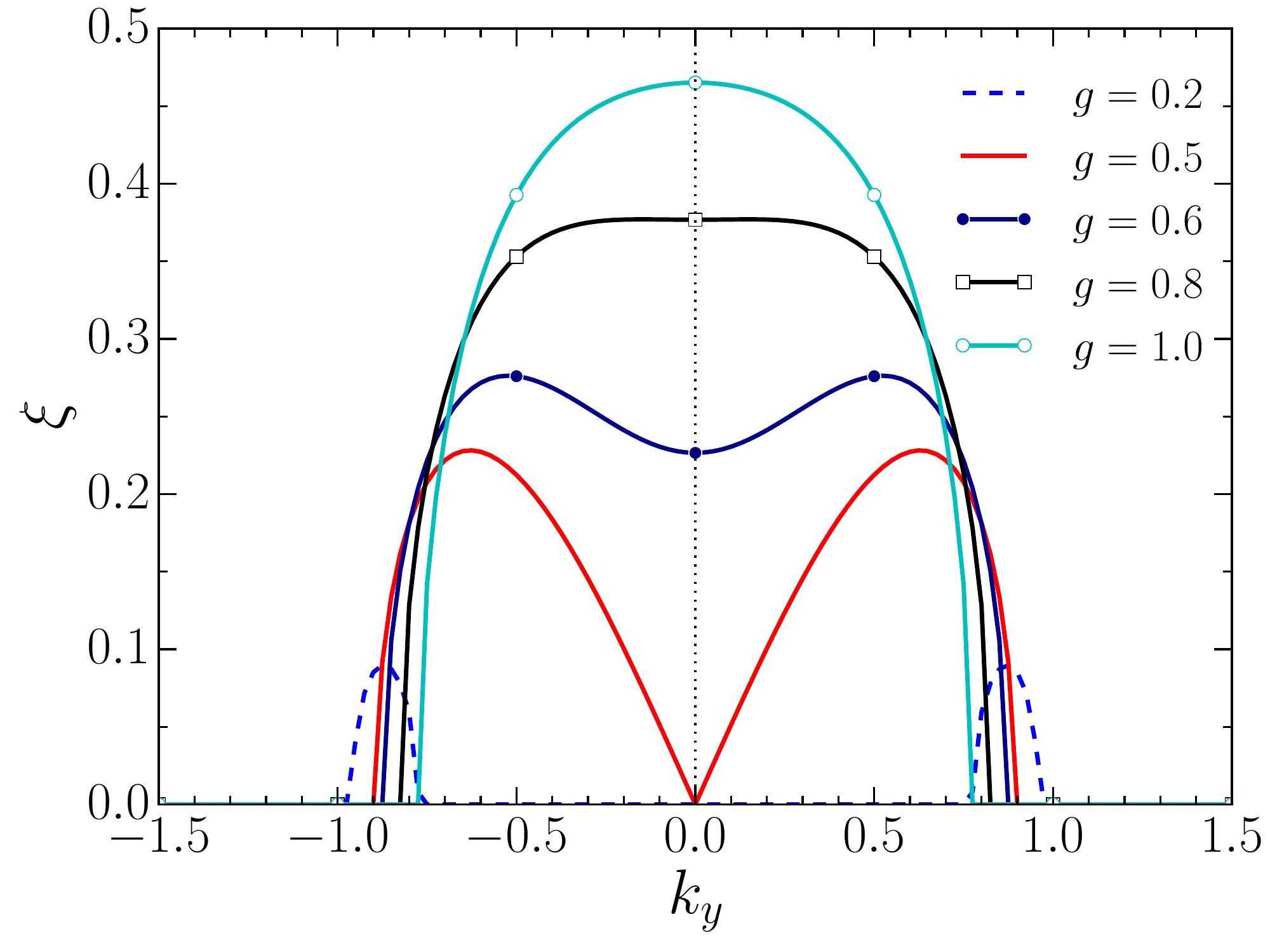}
\caption{(color online) Plot of the MI gain, $\xi = \lvert \text{Im} (\Omega _{+}) \rvert$, as a function of $k_y$ for different intra-component interaction strengths with $k_{L}= 1$, $\Gamma  =1$, $n=1$, $g_{12}=1$ and $k_x =1$.}
\label{fig-ls-2}
\end{figure}
Fig.~\ref{fig-ls-2} shows the MI gain for $\Omega_{+}$ as a function of one of the momentum component $(k_y)$ for different values of intra-component $(g)$ at fixed inter- component interaction strength $(g_{12}=1)$. The choice of parameters are $k_{L}=\Gamma  =1$, $n=1$, $g_{12}=1$ and $k_x =1$. It is evident from Fig.~\ref{fig-ls-2}, there exist two symmetrical instability region on either side of the zeros of $k_x$ and $k_y$.  As the intra-component interactions strength increases further, the two instability region approaches to the zero wave number and merge into a single coalesced instability region with elevated gain at higher values of~$g$.

\begin{figure}[!ht]
\centering\includegraphics[width=\linewidth]{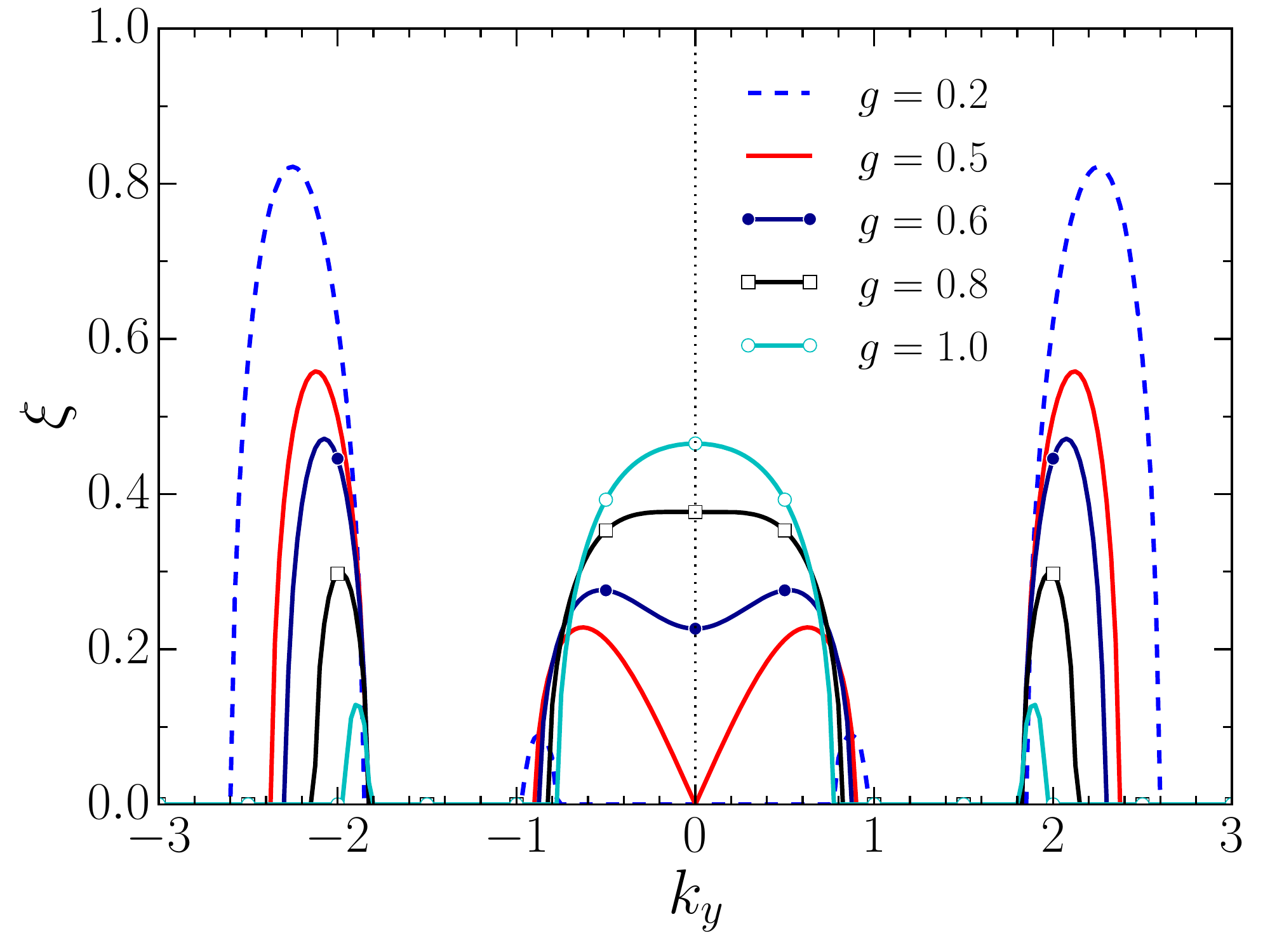}
\caption{(color online) Plot showing the MI gain, $\xi = \lvert \text{Im} (\Omega_{-}) \rvert$  as a function of $k_y$ for different $g$ with $k_{L}= 1$, $\Gamma  =1$, $n=1$, $g_{12}=1$ and $k_x =1$.}
\label{fig-ls-3}
\end{figure}

On the other hand, $\Omega_{-}$ from Eq.~(\ref{eq:om2d}) is more interesting, since $\Omega_{-}$ leads to unstable region even for $\varLambda^2 +4\Delta > 0$. Fig.~\ref{fig-ls-3} shows the MI gain for $\Omega_{-}$  as a function of momentum component for similar values as used for $\Omega_{+}$.
One can straightforwardly notice, that there exist two pairs of instability region for $\Omega_{-}$ as against, the single pair of instability region observed for the case $\Omega_{+}$. As the strength of the intra- component interaction increases, the instability region at the center unifies into single band (similar to the case of $\Omega_{+}$), while the other pair of instability region at higher values of $k_y$ substantially decreases in gain and width of the instability region.
\begin{figure}[!ht]%
\centering\includegraphics[width=\linewidth]{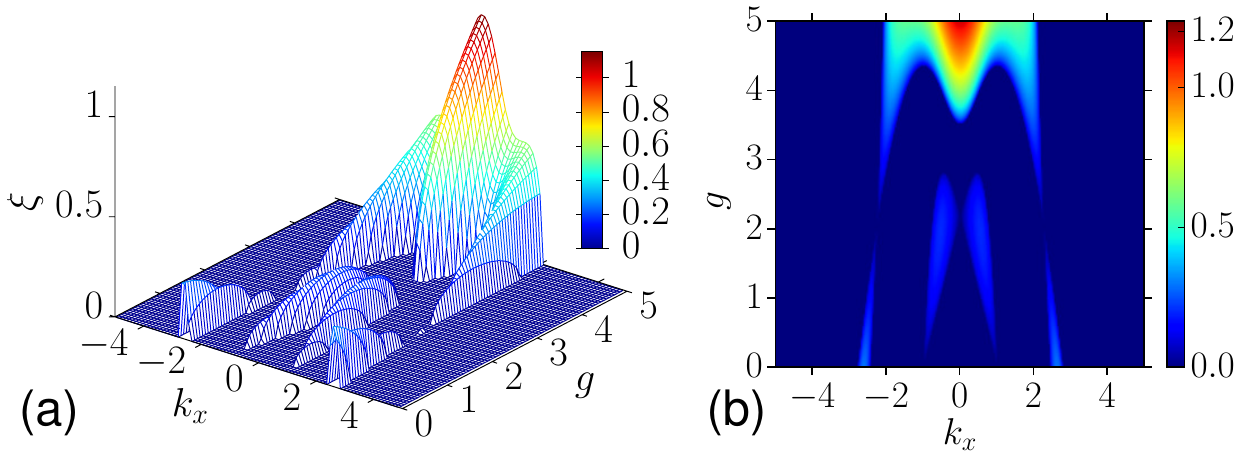}
\caption{ (color online) 3D surface plot of the MI gain, $\xi = \lvert \text{Im} (\Omega _{-}) \rvert$ in the $k_x$-$g$ plane and (b) the corresponding 2D contour plot for $k_{L}= 1$, $\Gamma  =1$, $n=0.3$, $g_{12} = 1$ and $k_y=1$.}
\label{fig-ls-4}
\end{figure}%
For insight, we plot in Fig.~\ref{fig-ls-4} the 3D variation of MI gain for a range of $k_x$ and $g$ with the inter-component interaction fixed $(g_{12})$. It is obvious that the two instability regions merge into a single instability region with elevated gain.
\begin{figure}[!ht]%
\centering\includegraphics[width=\linewidth]{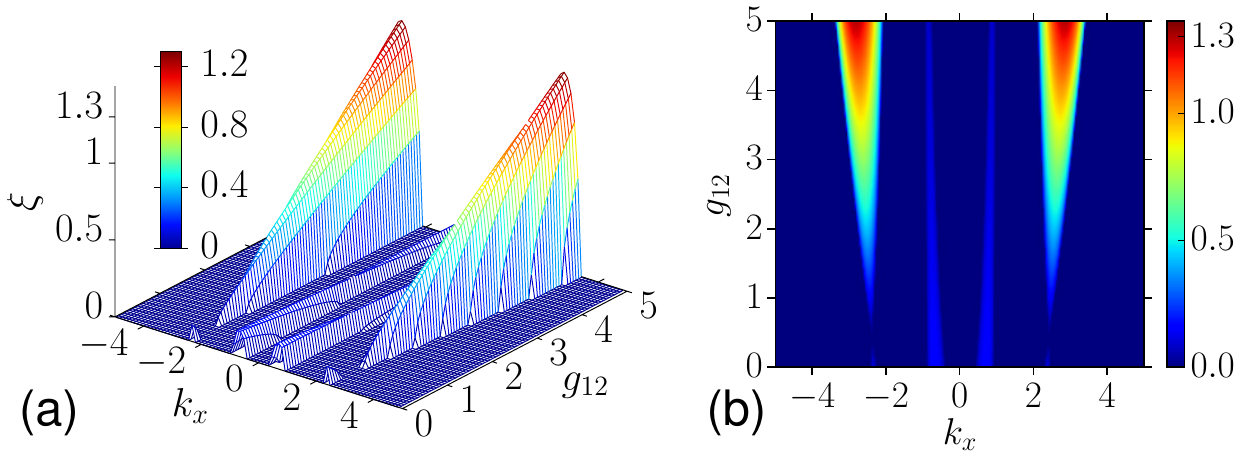}
\caption{ (color online) 3D surface plot showing the MI gain, $\xi = \lvert \text{Im} (\Omega _{-}) \rvert$, in the $k_x-g_{12}$ plane and (b) the corresponding 2D contour plot for the parameters $k_{L}=1$, $\Gamma =1$ for $n=0.3$, $g=1$ and $k_y=1$.}
\label{fig-ls-5}
\end{figure}%
To explore the effect of inter-component interaction in the instability, in Fig.~\ref{fig-ls-5}, we depict the MI gain for a range of $g_{12}$ with fixed $g$. It is apparent from Fig.~\ref{fig-ls-5}, for smaller values of~$g_{12}$, the gain in the inner instability band decreases gradually to zero, while the gain in the instability region at higher values of $k_x$ grows with increase in $g_{12}$.  In order to explore the effect of the wave numbers, $k_x$ and $k_y$, we plot the MI gain as a function of $k_x$ for different values of $k_y$ and vice-versa.
\begin{figure}[!ht]%
\centering\includegraphics[width=\linewidth]{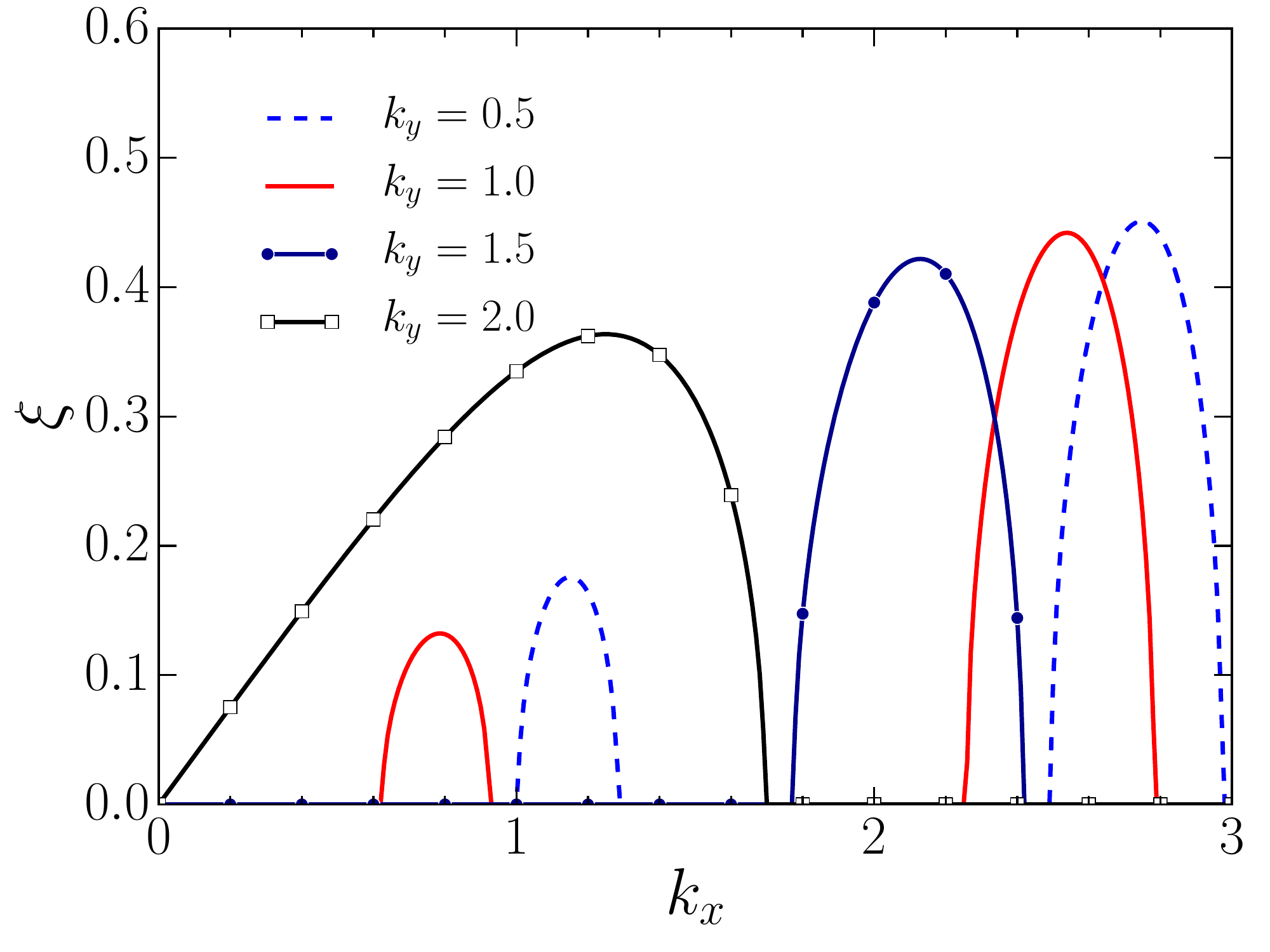}
\caption{(color online) Plot showing the MI gain, $\xi = \lvert \text{Im} (\Omega_{-}) \rvert$ as a function of $k_x$ for different $k_{y}$ values with $k_{L}= 1$, $\Gamma  =1$, $n=0.3$ $g=1$ and $g_{12}=2$.}
\label{kx}
\end{figure}%
\begin{figure}[!ht]%
\centering\includegraphics[width=\linewidth]{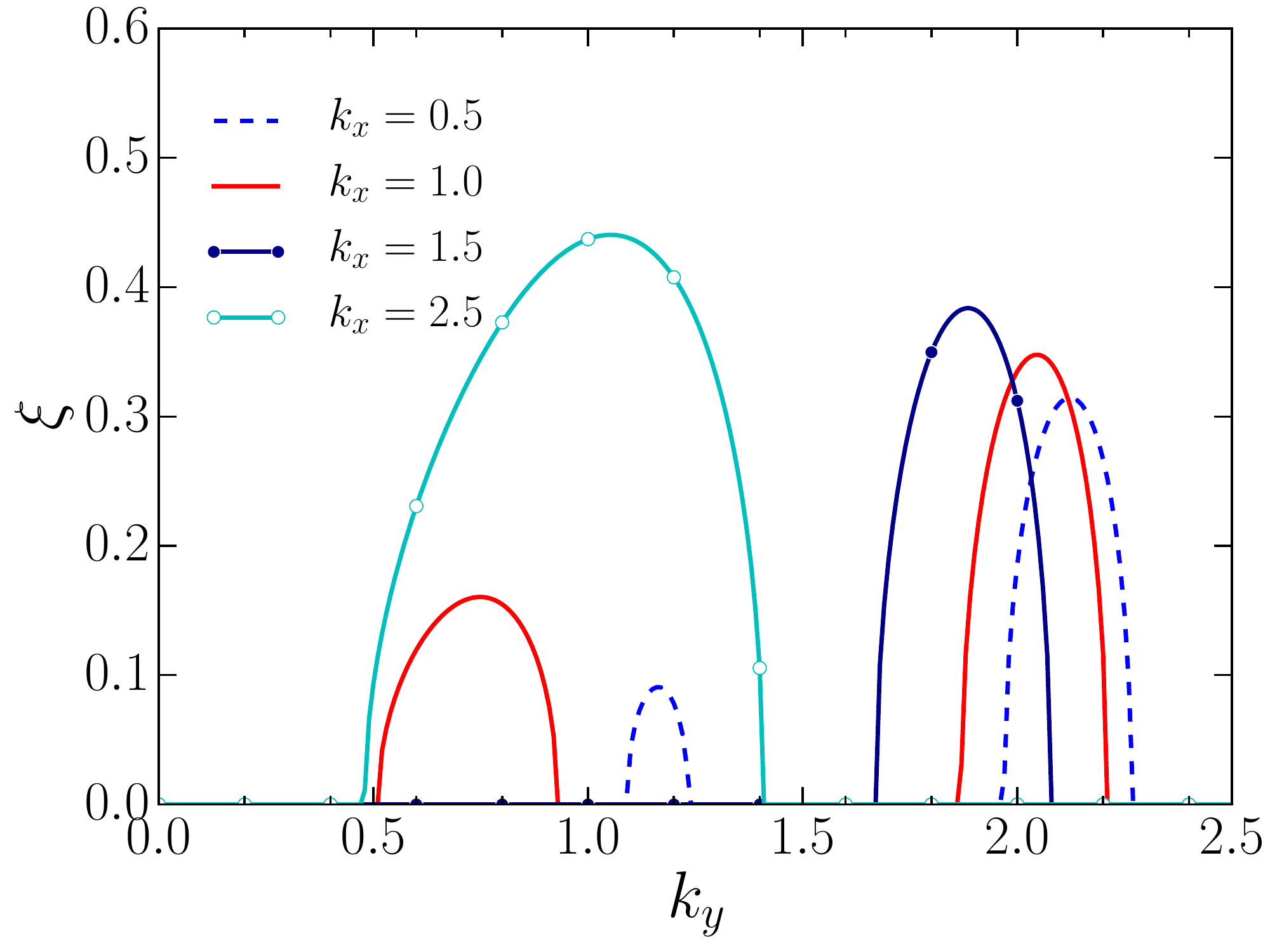}
\caption{(color online) Plot of the MI gain, $\xi = \lvert \text{Im} (\Omega_{-}) \rvert$ as a function of $k_y$ for different $k_{x}$ values with $k_{L}= 1$, $\Gamma  =1$, $n=0.3$, $g=1$ and $g_{12}=2$.}
\label{ky}
\end{figure}%
Figs.~\ref{kx} and \ref{ky} show that the MI bands drift towards the center and coalesced into single instability band with the increase in the wave numbers.  Thus, there are no changes in the general trend of shifting of MI band for both cases, however the peak gain and the width of instability region substantially differs. One can infer that the maximum gain is observed for $k_x$ as evident from Fig.~\ref{kx} in comparison to the plot of MI gain for $k_y$ in Fig.~\ref{ky}.
\begin{figure}[!ht]%
\centering\includegraphics[width=0.99\linewidth]{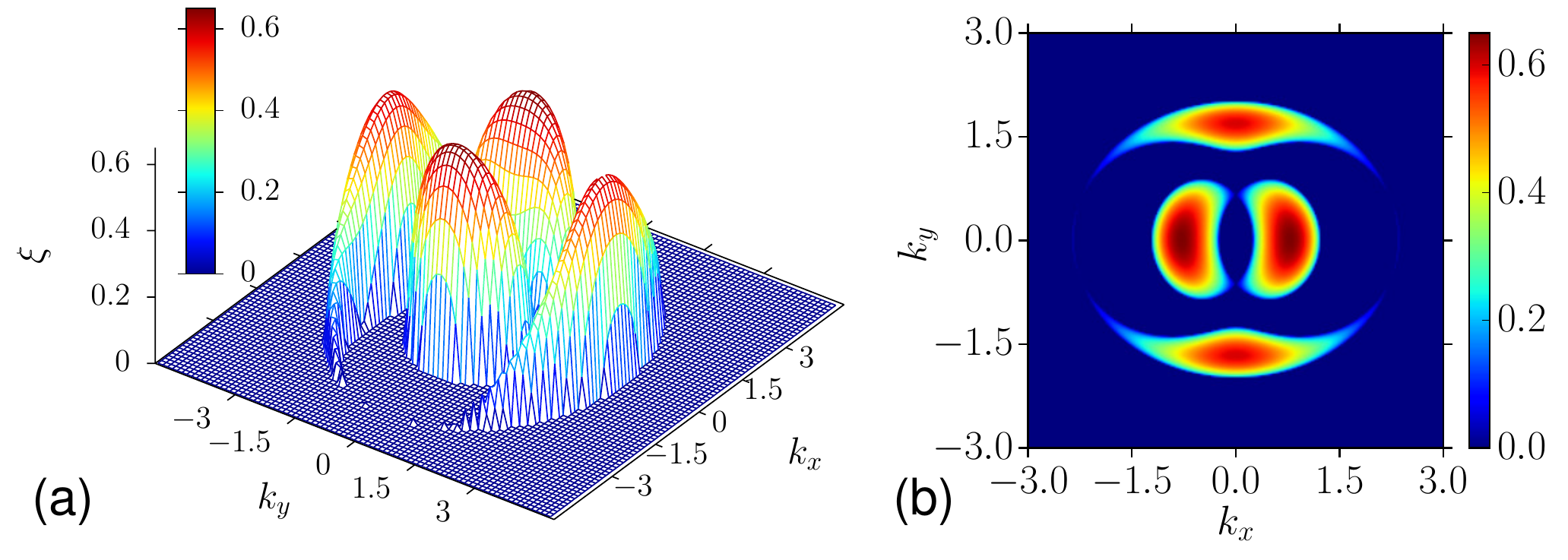}
\caption{(color online) (a) 3D surface plot showing the MI gain, $\xi = \lvert \text{Im} (\Omega _{-})\rvert$, and (b) the corresponding 2D contour plot for the parameters $k_{L}=1$, $\Gamma =1$, $n=0.3$, $g=5$ and $g_{12}=3$.}
\label{fig-ls-7-1}
\end{figure}%
Fig.~\ref{fig-ls-7-1} shows the instability gain on the momentum space as a function of $k_x$ and $k_y$ for some representative values of intra- and inter-components interaction strength.  It is observed that there exist two instability bands corresponding to $k_x$ and $k_y$. The inner pair of bands corresponds to $k_x$ with slightly higher gain than the outer band as a result of $k_y$.  This combination of intra- and inter-components interaction is of particular interest, because the instability is generally not feasible, as both interaction components are repulsive and therefore does not contribute to MI. However, the above results suggest that the MI is still possible even in the repulsive two component BEC with the aide of SOC.

\subsection{Attractive intra-component and repulsive inter-component interactions}
\label{sec:4B}
This condition corresponds to the binary BEC with attractive intra- component and repulsive inter-component interactions, which is subject to the MI even in the absence of the SOC. Although SOC is not fundamental to the occurrence of MI in this particular case, but significantly affects the instability.  The MI corresponding to $\Omega_{+}$ produces the same number of bands as in the previous case for repulsive interaction. 
\begin{figure}[!ht]
\centering\includegraphics[width=0.99\linewidth]{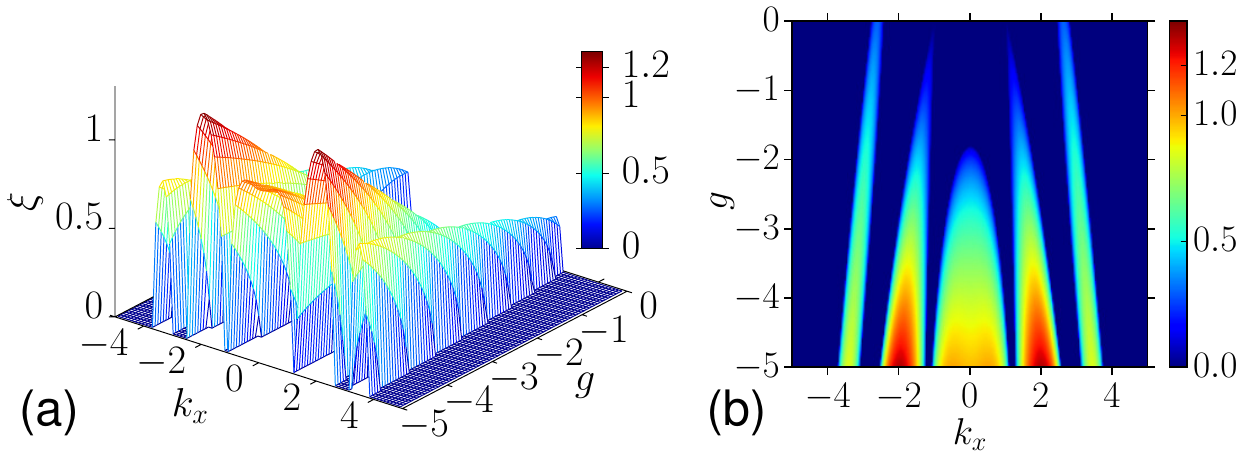}
\caption{ (color online)
3D surface plot of the MI gain, $\xi = \lvert \text{Im} (\Omega _{-}) \rvert$ in the $k_x$-$g$ plane and (b) the corresponding 2D contour plot for fixed $g_{12}$ with  $k_{L}= 1$, $\Gamma  =1$, $n=0.3$, $g_{12} = 1$ and $k_y=1$.}
\label{fig-ls-9}
\end{figure}
However, the MI corresponding to $\Omega_{-}$  qualitatively differs, and it can be better explained in the following two combinations, namely, (i) MI gain as a function of $g$ for fixed $g_{12}$, and (ii) variation of MI gain as a function of $g_{12}$ at constant value of~$g$. Fig.~\ref{fig-ls-9} shows the possibility of three pairs of instability bands for attractive intra-component at constant~$g_{12}$, and the instability bands grow in gain with the increase in~$g$. 
\begin{figure}[!ht]
\centering\includegraphics[width=0.99\linewidth]{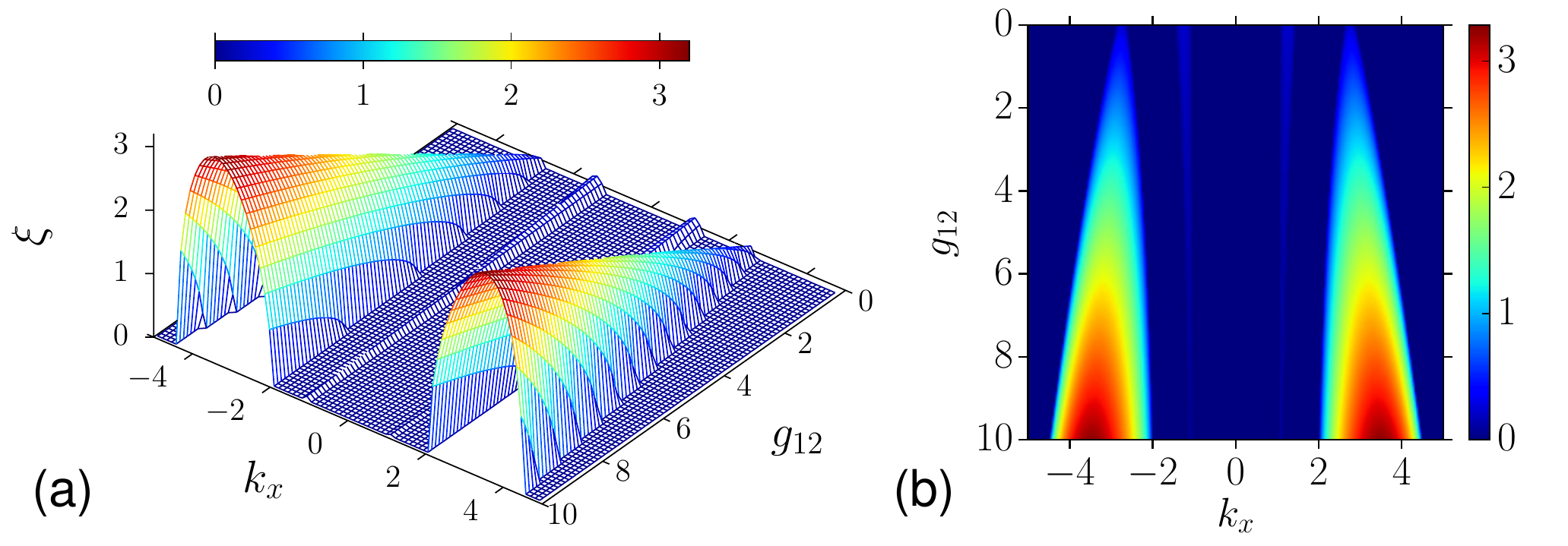}
\caption{ (color online)
3D surface plot of the MI gain, $\xi = \lvert \text{Im} (\Omega _{-}) \rvert$ in the $k_x$-$g_{12}$ plane and (b) the corresponding 2D contour plot for fixed $g$ with $k_{L}= 1$, $\Gamma  =1$, $n=0.3$, $g=-1$ and $k_y=1$.}
\label{fig-ls-6}
\end{figure}
Fig.~\ref{fig-ls-6} depicts the MI gain for a range of repulsive inter-component interaction strength ($g_{12}$) at constant~$g$. It is obvious, there exist two pairs of instability bands on either side of $k_y$, the inner one decreases with increase in $g_{12}$, while the outer own grow with the increase in $g_{12}$. The instability gain in momentum space for some representative value of intra- and inter-components interaction strength is shown in Fig.~\ref{fig-ls-4-1}. 
\begin{figure}[!ht]%
\centering\includegraphics[width=0.99\linewidth]{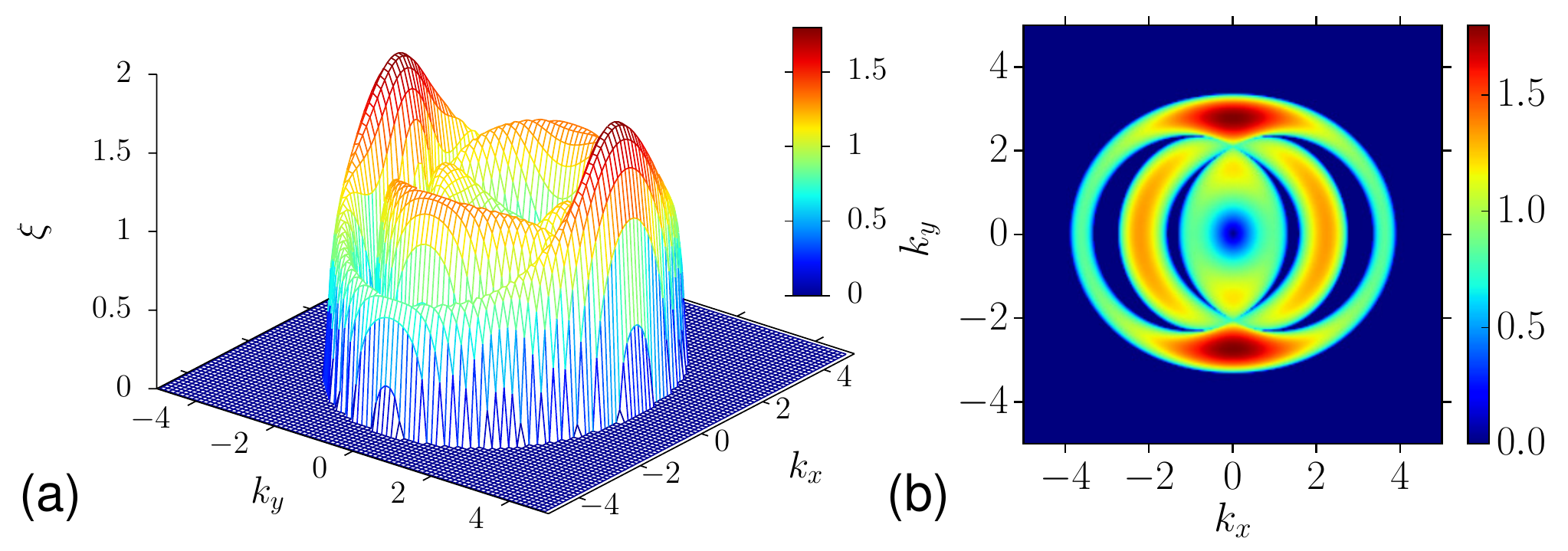}
\caption{(color online) (a) 3D surface plot showing the MI gain, $\xi = \lvert \text{Im} (\Omega _{-})\rvert$, and (b) the corresponding 2D contour plot for the parameters $k_{L}=1$, $\Gamma =1$, $n=0.3$, $g=-5$ and $g_{12}=1$.}
\label{fig-ls-4-1}
\end{figure}%
It is observed that there exist three symmetric instability bands corresponding to $k_x$, while only two for $k_y$. Unlike the previous case, the instability gain is maximum for the bands corresponding to $k_y$ as shown in Fig.~\ref{fig-ls-4-1}.

\subsection{Repulsive intra-component and attractive inter-component interactions}
\label{sec:4C}
Here, we consider the binary BEC with repulsive intra-component $(g>0)$ and attractive inter-component interaction $g_{12}<0$. It is obvious from our earlier discussion, in the absence of Rabi and SO coupling, the MI (through $\Omega_{-}$) is observed provided the condition $\lvert g_{12} \rvert >\lvert g \rvert$ is satisfied. But in the presence of SO coupling, the MI is said to occur regardless of the sign of the interaction strength and there are no conditions imposed. In similar lines with the previous section, we discuss MI in the two particular cases, i.e. constant intra-component interaction strength with varying inter-component strength and vice-versa.
\begin{figure}[!ht]%
\centering\includegraphics[width=0.99\linewidth]{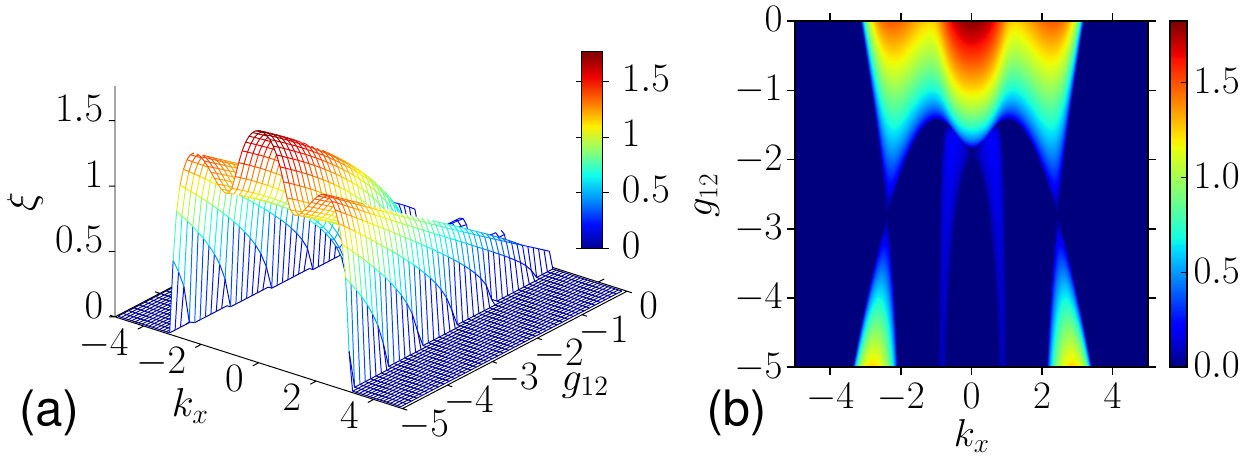}
\caption{ (color online)
3D surface plot of the MI gain, $\xi = \lvert \text{Im} (\Omega _{-}) \rvert$ in the $k_x$-$g_{12}$ plane and (b) the corresponding 2D contour plot for fixed $g$ with $k_{L}= 1$, $\Gamma  =1$, $n=0.3$, $g=1$ and $k_y=1$.}
\label{Fig5-1}
\end{figure}%
Fig.~\ref{Fig5-1} shows the MI gain at constant $g$ as a function of $g_{12}$. As the strength of the inter-component interaction increases, the outer instability band grows and merges with inner instability band of higher gain.
\begin{figure}[!ht]%
\centering\includegraphics[width=0.99\linewidth]{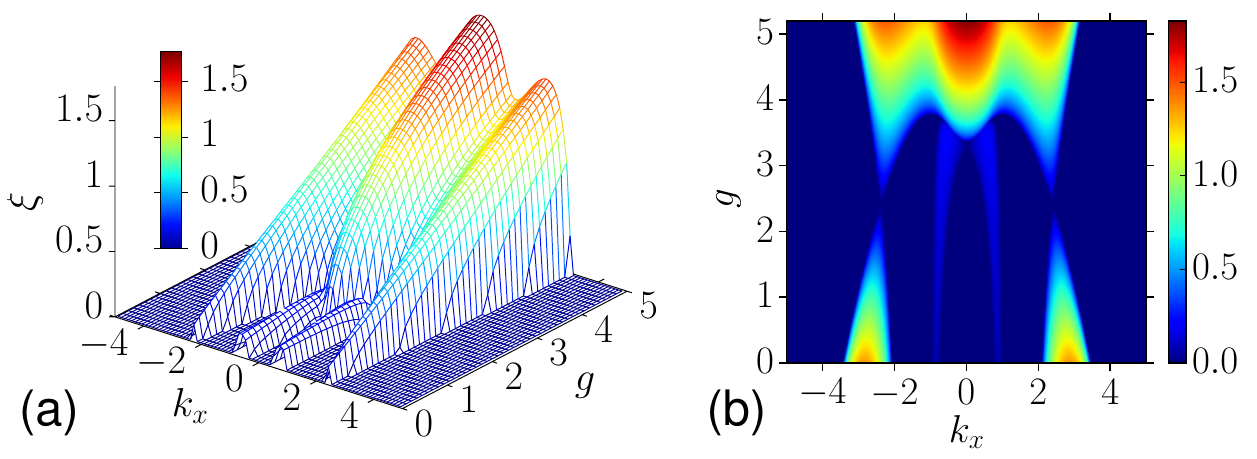}
\caption{ (color online)
3D surface plot of the MI gain, $\xi = \lvert \text{Im} (\Omega _{-}) \rvert$ in the $k_x$-$g$ plane and (b) the corresponding 2D contour plot for fixed $g_{12}$ with $k_{L}= 1$, $\Gamma  =1$, $n=0.3$, $g_{12}=-1$ and $k_y=1$.}
\label{Fig5-2}
\end{figure}%
The variation of MI gain for $g_{12}$ at constant $g$ shows a similar trend, except the changes in the numerical value of gain as shown in Fig.~\ref{Fig5-2}. Fig.~\ref{fig-ls-5-1} depicts the MI gain in momentum space for $k_x$ and $k_y$. 
\begin{figure}[!ht]%
\centering\includegraphics[width=0.99\linewidth]{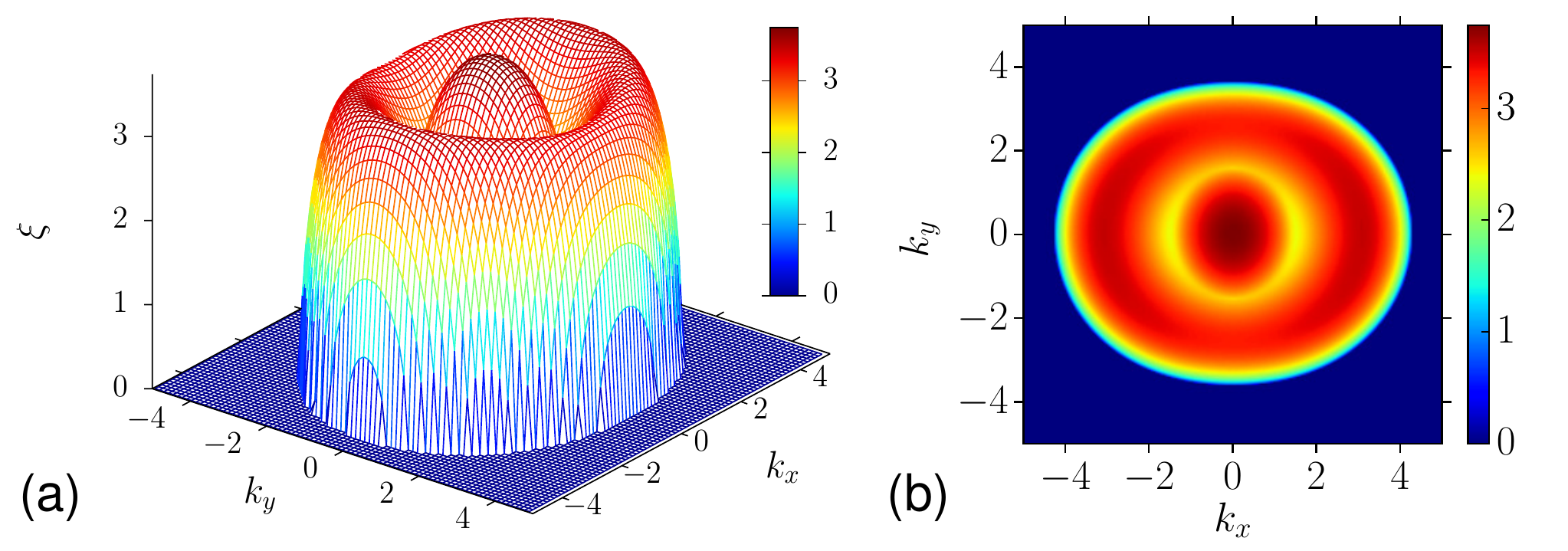}
\caption{(color online) (a) 3D surface plot showing the MI gain, $\xi = \lvert \text{Im} (\Omega _{-})\rvert$, and (b) the corresponding 2D contour plot for the parameters $k_{L}=1$, $\Gamma =1$, $n=0.3$, $g=2$ and $g_{12}=-13$.}
\label{fig-ls-5-1}
\end{figure}%
Unlike the earlier cases, the gain of the inner band is quantitatively same for both bands corresponding to $k_x$ and $k_y$. However, the instability gain of the outer band corresponding to $k_y$ is slightly larger than the outer band of $k_y$.

\subsection{Attractive intra-and inter-component interactions}
\label{sec:4D}
In this region, both intra- and inter-component interactions are attractive, i.e. $g<0$ and $g_{12}<0$, and therefore, MI occurs naturally even without the aide of Rabi and SOC. This case has already been discussed thoroughly in the context of MI in two component BEC, and hence, an extensive investigation is needless. However, for the sake of completeness, we focus on the effect of SOC in the instability.  
\begin{figure}[!ht]%
\centering\includegraphics[width=0.99\linewidth]{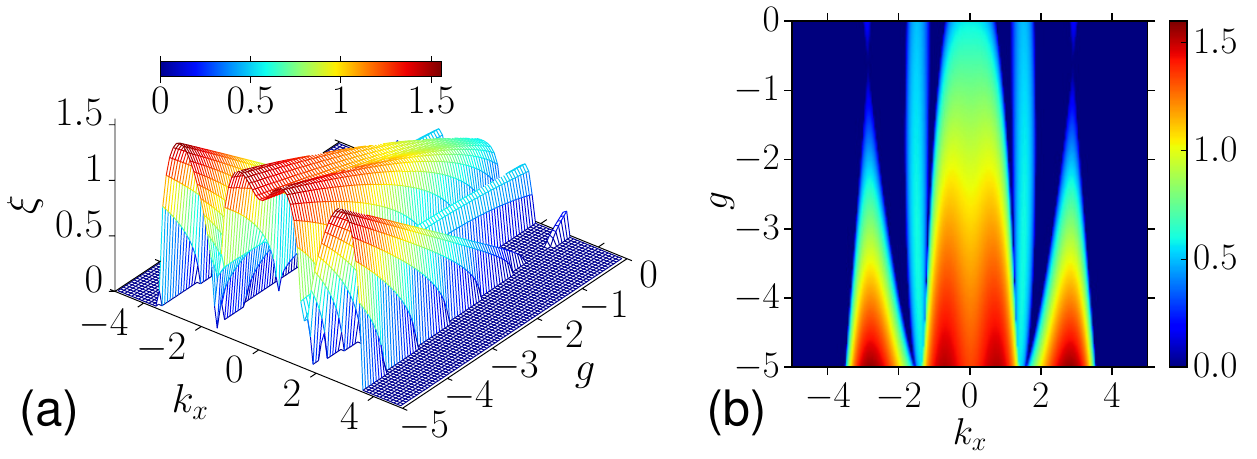}
\caption{ (color online)
3D surface plot of the MI gain, $\xi = \lvert \text{Im} (\Omega _{-}) \rvert$ in the $k_x$-$g$ plane and (b) the 2D  corresponding contour plot for fixed $g_{12}$ with $k_{L}= 1$, $\Gamma  =1$,  $n=0.3$, $g_{12}=-1$, and $k_y=1$.}
\label{Fig15}
\end{figure}%
Fig.~\ref{Fig15} shows that the growth of MI gain with the variation of $g_{12}$ for constant $g=-1$. The current case completely concur with our earlier discussion, and the MI becomes independent of the $g$ for the strong $g_{12}$ interaction. The variation of MI gain in momentum space is shown in Fig.~\ref{fig-ls-6-1}. 
\begin{figure}[!ht]%
\centering\includegraphics[width=0.99\linewidth]{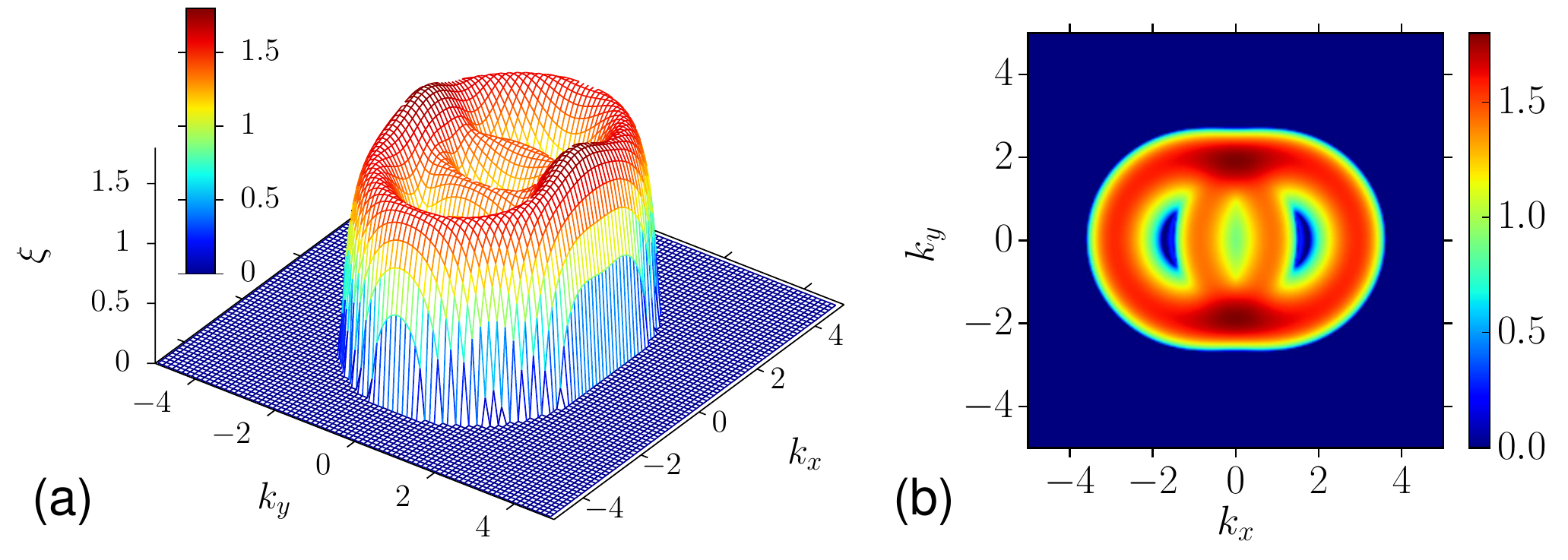}
\caption{(color online) (a) 3D surface plot showing the MI gain, $\xi = \lvert \text{Im} (\Omega _{-})\rvert$, and (b) the corresponding 2D contour plot for the parameters $k_{L}=1$, $\Gamma =1$, $n=0.3$, $g=-1$, $g_{12}=-5$.}
\label{fig-ls-6-1}
\end{figure}%
Like in the previous section, the MI bands are symmetric across the zero wave number, and the maximum gain occurs for the bands corresponding to $k_y$.

\begin{table*}[ht!]
\caption {Summary of MI in SO coupled two dimensional binary BEC}
\label{table1}
\begin{center}
\begin{tabular}{ |l|l|l|l|l| }
\hline
\,\,\textbf{SO coupling}\,\, &\,\, \textbf{Rabi coupling} \,\, &\,\,\textbf{ MI gain} \,\, &\,\, \textbf{Different combinations} \,\, &\,\, \textbf{Inference}\,\, \\
\hline
\multirow{4}{*}{} & & &\,\, both interaction are repulsive \,\, & \,\, Always stable \,\,\\
& & & \,\, $g<0$, $g_{12}>0$ & \,\, Always stable\\
&  & \,\,\,\,\,\,\,\, $\Omega _{+}$ \,\,\,\,\,\,\,\, & \,\, $g>0$, $g_{12}<0$ \,\,& \,\, Always stable\\
& & & \,\, both interaction are attractive \,\, & \,\, Always unstable \,\,\\  \cline{3-5}

\multirow{4}{*}{} & \,\,\,\,\,\,\,\,\,\,\,\, $\Gamma =0$ \,\,\,\,\,\,\,\,\,\,\,\, & \,\, & \,\, both interaction are repulsive & \,\,  \\
&  & & \,\, $g<0$, $g_{12}>0$  & \,\, stable for all cases\\
& \,\, & \,\,\,\,\,\,\,\, $\Omega _{-}$ \,\,\,\,\,\,\,\, & \,\,  $g>0$, $g_{12}<0$ & \,\,   \\
& & & \,\, both interaction are attractive \,\,& \,\, \\  \cline{2-5}

\multirow{4}{*}{} & & & \,\, both interaction are repulsive & \,\, \\
& & & \,\, $g<0$, $g_{12}>0$  &\,\, Similar to the case \\
& & \,\,\,\,\,\,\,\, $\Omega _{+}$ \,\,\,\,\,\,\,\, & \,\, $g>0$, $g_{12}<0$ & \,\, with $\Gamma =0$ and $\Omega _{+}$ \\
&  & & \,\, both interaction are attractive & \,\, \\  \cline{3-5}

\multirow{4}{*} \,\,\,\,\,\,\,\, {$k_{L}=0$} \,\,\,\,\,\,\,\, & \,\,\,\,\,\,\,\,\,\,\,\, $\Gamma <0$ \,\,\,\,\,\,\,\,\,\,\,\, & \,\, & \,\, both interaction are repulsive & \,\, Always stable \\
&  & & \,\, $g<0$, $g_{12}>0$ & \,\, Always unstable \\
& & \,\,\,\,\,\,\,\, $\Omega _{-}$ \,\,\,\,\,\,\,\, & \,\, $g>0$, $g_{12}<0$ & \,\, Always stable \\
&  & & \,\, both interaction are attractive & \,\, Always stable \\  \cline{2-5}

\multirow{4}{*}{} & & & \,\, both interaction are repulsive & \,\, \\
& & & \,\, $g<0$, $g_{12}>0$  &\,\, Similar to the case \\
& & \,\,\,\,\,\,\,\, $\Omega _{+}$ \,\,\,\,\,\,\,\, & \,\, $g>0$, $g_{12}<0$ & \,\, with $\Gamma =0$ and $\Omega _{+}$ \\
&  & & \,\, both interaction are attractive & \,\, \\  \cline{3-5}

\multirow{4}{*}{} & \,\,\,\,\,\,\,\,\,\,\,\, $\Gamma >0$ \,\,\,\,\,\,\,\,\,\,\,\, & \,\, & \,\, both interaction are repulsive & \,\, Unstable if $\lvert g \rvert >\lvert g_{12} \rvert$  \\
&  & & \,\, $g<0$, $g_{12}>0$ & \,\, Always stable \\
&  & \,\,\,\,\,\,\,\, $\Omega _{-}$ \,\,\,\,\,\,\,\, & \,\, $g>0$, $g_{12}<0$ & \,\, Always unstable\\
&  & & \,\, both interaction are attractive & \,\, Unstable if $\lvert g_{12} \rvert >\lvert g \rvert$\\
\hline

\multirow{4}{*}{} & & & \,\, both interaction are repulsive & \,\, Always stable \\
&  & & \,\, $g<0$, $g_{12}>0$ & \,\, Unstable  if $\lvert g_{12} \rvert >\lvert g \rvert$ \\
&  & \,\,\,\,\,\,\,\, $\Omega _{+}$ \,\,\,\,\,\,\,\, & \,\, $g>0$, $g_{12}<0$ \,\, & \,\, Always stable\\
&  & & \,\, both interaction are attractive & \,\, Unstable if $\lvert g \rvert >\lvert g_{12} \rvert$ \\ \cline{3-5}

\multirow{4}{*}{} & \,\,\,\,\,\,\,\,\,\,\,\, $\Gamma =0$ \,\,\,\,\,\,\,\,\,\,\,\, &\,\, &\,\, both interaction are repulsive & \,\, Unstable if $\lvert g_{12} \rvert >\lvert g \rvert$ \\
&  & & \,\, $g<0$, $g_{12}>0$ & \,\, Always unstable \\
& & \,\,\,\,\,\,\,\, $\Omega _{-}$ \,\,\,\,\,\,\,\, & \,\, $g>0$, $g_{12}<0$ & \,\, Unstable if $\lvert g \rvert >\lvert g_{12} \rvert$ \\
& & & \,\, both interaction are attractive \,\, & \,\, Always unstable \\  \cline{2-5}

\multirow{4}{*}{} &  & &\,\, both interaction are repulsive & \,\,  \\
&  & & \,\, $g<0$, $g_{12}>0$ & \,\,  stable for all  cases\\
&  & \,\,\,\,\,\,\,\, $\Omega _{+}$ \,\,\,\,\,\,\,\, & \,\, $g>0$, $g_{12}<0$ & \\
&  & & \,\, both interaction are attractive & \,\,  \\  \cline{3-5}

\multirow{4}{*} \,\,\,\,\,\,\,\, {$k_{L}=1$} \,\,\,\,\,\,\,\, & \,\,\,\,\,\,\,\,\,\,\,\, $\Gamma <0$ \,\,\,\,\,\,\,\,\,\,\,\, & \,\, & \,\, both interaction are repulsive & \,\, Always stable \\
&  & & \,\, $g<0$, $g_{12}>0$ & \,\, Always unstable \\
& & \,\,\,\,\,\,\,\, $\Omega _{-}$ \,\,\,\,\,\,\,\, & \,\, $g>0$, $g_{12}<0$ & \,\, Unstable if $\lvert g \rvert >\lvert g_{12} \rvert$  \\
&  & & \,\, both interaction are attractive & \,\, Always unstable \\  \cline{2-5}

\multirow{4}{*}{} & & & \,\, both interaction are repulsive & \,\, Always unstable \\
& & & \,\, $g<0$, $g_{12}>0$ & \,\, Always unstable \\
& & \,\,\,\,\,\,\,\, $\Omega _{+}$ \,\,\,\,\,\,\,\, & \,\, $g>0$, $g_{12}<0$ \,\, & \,\, Unstable  if $\lvert g \rvert >\lvert g_{12} \rvert$  \\
&  & & \,\, both interaction are attractive & \,\, Always unstable \\  \cline{3-5}

\multirow{4}{*}{} & \,\,\,\,\,\,\,\,\,\,\,\, $\Gamma >0$ \,\,\,\,\,\,\,\,\,\,\,\, & \,\, & \,\, both interaction are repulsive & \,\,  \\
&  & & \,\, $g<0$, $g_{12}>0$  & \,\, Always unstable \\
&  & \,\,\,\,\,\,\,\, $\Omega _{-}$ \,\,\,\,\,\,\,\, & \,\, $g>0$, $g_{12}<0$ & \,\,  \\
&  & & \,\, both interaction are attractive & \,\,  \\
\hline
\end{tabular}
\end{center}
\end{table*}

\subsection{Results and Discussion}

For the ease of understanding and to make the analysis self-explanatory, we summarize our results of MI in the two-dimensional SO coupled two-component BEC in Table~\ref{table1}.  We systematically discussed the presence/absence of SO/Rabi coupling under different combination of signs of intra- and inter-component interaction strength. As it is evident from our extensive investigation that SO coupling inevitably destabilizes the initial steady state for equal densities of binary BEC, and thereby makes the system unstable for all combinations of interaction strength. Also, we have shown the conventional MI immiscibility condition, $g_{12}>g$ for repulsive two component BEC system is no longer significant for MI. Our particular focus is on repulsive intra- and inter-component interaction, as it is proven to be stable against the perturbation, and therefore, MI is generally not feasible.  However, we have shown that MI can be achieved with the effect of SOC as demonstrated through Figs.~\ref{fig-ls-2} - \ref{fig-ls-7-1}. We discussed the MI gain in momentum space as a function of $k_x$ and $k_y$ and emphasize the variation of gain over the wave numbers in the two directions. We noted that the MI gain is not identical on $k_x$ and $k_y$, and significant changes in MI gain, width of instability region and the number of instability bands are readily observed.

In Sec.~\ref{sec:4B}, we discussed the MI condition for attractive intra- and repulsive inter-component interactions. Figs.~\ref{fig-ls-9} - \ref{fig-ls-4-1} show the instability gain as a function of $g$ and $g_{12}$. One can straightforwardly observe the emergence of new instability bands in the MI gain plot, which is identified to be the consequence of the incorporation of SOC. Following that, we discussed in Sec. \ref{sec:4C}, the MI scenario in the case of repulsive intra- and attractive inter-component interactions.  Figs.~\ref{Fig5-1} - \ref{fig-ls-5-1} portray the variation of MI gain for $g$ and $g_{12}$. Along the similar lines with the earlier cases, the SOC results in new instability bands and thereby help in enhancing the MI in such systems. Finally, we studied MI in attractive intra- and inter-component interactions in Sec. \ref{sec:4D}.  It is very well known from the theory of MI in BEC, that attractive interactions naturally support MI. Although,  SOC is not fundamental for the origin of MI, however, SOC significantly influences the instability region in terms of peak gain and width as evident from Figs. \ref{Fig15} and \ref{fig-ls-6-1}. Overall, the effect of SOC can be understood as a means to achieve MI in repulsive interactions, and also enhance instability in the system.

Last but not least, it is also important to see the impact of SO and Rabi couplings on MI gain for fixed wavenumbers $k_x$ and $k_y$. 
\begin{figure}[!ht]%
\centering\includegraphics[width=0.99\linewidth]{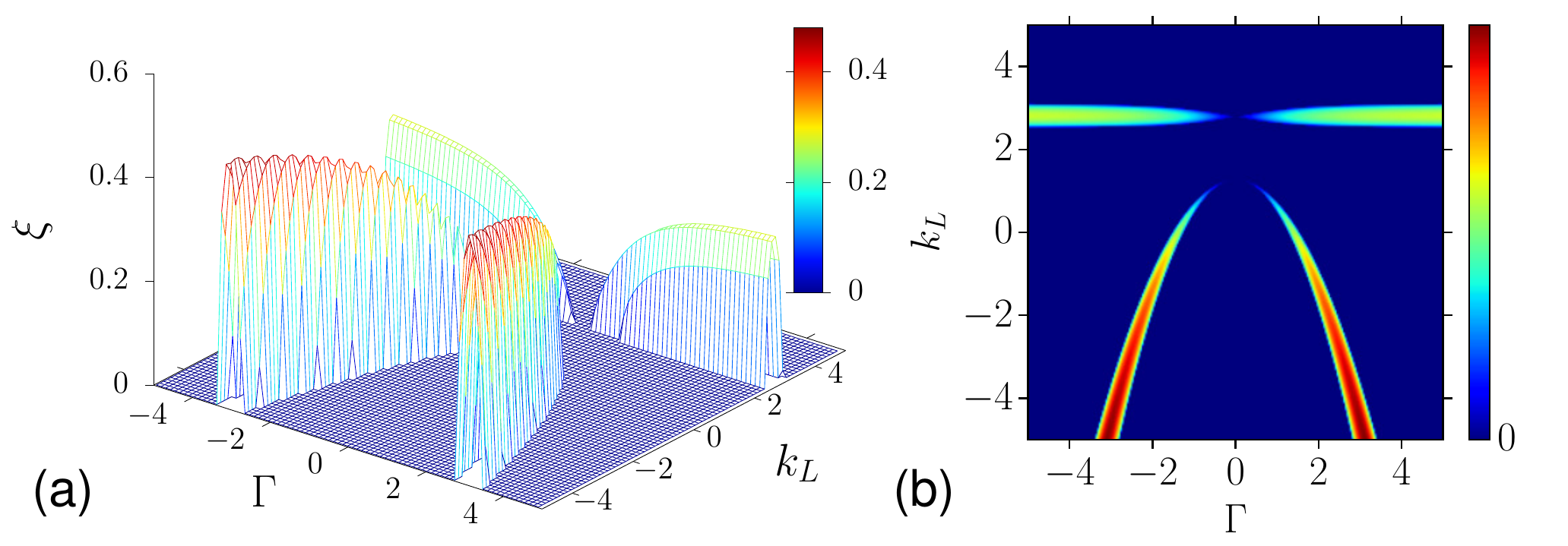}
\caption{(color online) (a) 3D surface plot showing the MI gain, $\xi = \lvert \text{Im} (\Omega _{-})\rvert$, and (b) the corresponding 2D contour plot for the parameters $k_{L}= 1$, $\Gamma  =1$, $g=1$, $g_{12}=1$, $n=0.3$, $k_{x}=2$ and $k_{y}=1$.}
\label{fig-ls-11}
\end{figure}%
Fig.~\ref{fig-ls-11} depicts the MI gain as a function of SO and Rabi coupling. As it is evident from our choice of parameters, the instability bands are symmetric for both positive and negative values of the Rabi and SO coupling. One can also infer, that the MI is possible even for zero SOC, provided the Rabi coupling is $\Gamma>0$.
\section{Conclusions}
\label{sec:5}

To summarize, we investigated the dynamics of MI gain in two-dimensional SO coupled binary BEC at an equal density of pseudo-spin components. The dispersion relation corresponding to the instability of the flat CW background against small perturbation was studied using linear stability analysis. For a comprehensive study, we consider all the possible combination of signs of intra- and inter-component interactions, with a particular, emphasize on repulsive interactions. Our analysis illustrates that SOC inevitably contributes to instability, regardless of the nature of the interaction strength.  With detailed interpretation, we have shown that the repulsive intra- and inter-component interaction admit instability and the MI immiscibility condition $g_{12}>g$ is no longer essential for MI. We also have shown, for the strong attractive inter-component interaction, the nature of the intra- component interaction is immaterial for constant SO and Rabi coupling.  We also analyzed the variation of instability domain in momentum space for $k_x$ and $k_y$.  The MI gain is not identical on $k_x$ and $k_y$, and significant changes in MI gain and a number of bands are observed. In the case of systems naturally admitting MI (attractive interactions), the SOC and Rabi coupling manifest in the generation of new instability bands, thereby enhances the MI. Thus, we presented a comprehensive analysis with detailed interpretation and graphical illustration of MI in two-dimensional SO coupled binary BEC for equal densities. We believe, the aforementioned results could potentially provide new ways to generate and manipulate MI and solitons in two-dimensional BECs.

\acknowledgments
The work of PM forms a part of Science \& Engineering Research Board, Department of Science \& Technology, Government of India sponsored research project (No. EMR/2014/000644). KP thanks the DST, IFCPAR, NBHM, and CSIR, Government of India, for the financial support
through major projects.

\end{document}